# Electrostatics of metal-graphene interfaces: sharp *p-n* junctions for electron-optical applications

Ferney A. Chaves[1], David Jiménez[1], Jaime E. Santos[2], Peter Bøggild[3], José M. Caridad[3†]

[1]*Department d'Enginyeria Electrònica, Escola d'Enginyeria, Campus UAB, Bellaterra, 08193 Barcelona, Spain*

[2]*Centro de Física and Departamento de Física, Universidade do Minho, P-4710-057 Braga, Portugal*

[3]*Center for Nanostructured Graphene (CNG), Department of Physics, Technical University of Denmark, 2800 Kongens Lyngby, Denmark*

[†]corresponding author: jcar@dtu.dk

**ABSTRACT.**

**Creation of sharp lateral *p-n* junctions in graphene devices, with transition widths *w* well below the Fermi wavelength $\lambda_F$ of graphene's charge carriers, is vital to study and exploit these electronic systems for electron-optical applications. The achievement of such junctions is, however, not trivial due to the presence of a considerable out-of-plane electric field in lateral *p-n* junctions, resulting in large widths. Metal-graphene interfaces represent a novel, promising and easy to implement technique to engineer such sharp lateral *p-n* junctions in graphene field-effect devices, in clear contrast to the much wider (i.e. smooth) junctions achieved via conventional local gating. In this work, we present a systematic and robust investigation of the electrostatic problem of metal-induced lateral *p-n* junctions in gated**

**graphene devices for electron-optics applications, systems where the width *w* of the created junctions is not only determined by the metal used but also depends on external factors such as device geometries, dielectric environment and different operational parameters such as carrier density and temperature. Our calculations demonstrate that sharp junctions ($w \ll \lambda_F$) can be achieved via metal-graphene interfaces at room temperature in devices surrounded by dielectric media with low relative permittivity (<10). In addition, we show how specific details such as the separation distance between metal and graphene and the permittivity of the gap in-between plays a critical role when defining the p-n junction, not only defining its width *w* but also the energy shift of graphene underneath the metal. These results can be extended to any two-dimensional (2D) electronic system doped by the presence of metal clusters and thus are relevant for understanding interfaces between metals and other 2D materials.**

1. Introduction.

*Sharp lateral p-n junctions in graphene for electron optics applications.*

Electronic *p-n* junctions are well-known, fundamental constituents of current semiconductor technology, enabling, for instance, the control of current flow passing through devices such as diodes and transistors or allowing photoelectric conversion in solar cells. [1] These junctions are formed at the interface between two differently doped semiconductor regions or materials, one with excess of holes (*p*-type doped) and one with excess of electrons (*n*-type doped), where a built-in potential gradient across such *p-n* boundary is the actual responsible for the rectification of currents or the separation of photo-generated electron-hole pairs. The advent of semiconducting two-dimensional (2D) crystals has revived interest in studying *p-n* junctions from both fundamental and an applications' point of view. [2-4]. To a large extent, this is due to the unique possibility of creating lateral, in-plane *p-n* junctions of similar or different polarity (*n-n'*, *n-p*, *p-n* and *p-p'*) in these semiconducting 2D systems [3-7]: being formed at the one-dimensional interface between two coplanar 2D regions, lateral *p-n* junctions do not have a 3D counterpart. In practical terms, these *p-n* junctions offer attractive opportunities to realize diodes with large rectification ratios that can be modulated by a gate-voltage [8] or photodetectors with high responsivity [9]. Moreover, lateral *p-n* junctions in ballistic 2D electron systems enable the creation of novel and exciting device concepts based on the possibility to steer the electron flow at these interfaces analogously to the way optical lenses and prisms direct light in an optical apparatus.[10-15]

Among all potential candidates, graphene is arguably the most alluring 2D solid-state system to study and implement electron optics devices based on lateral *p-n* junctions (Fig. 1a) due to several reasons. First, it displays ballistic transport over micrometer length scales even at room temperature [16,17], relevant for applications. In addition, being a zero-gap material with a linear dispersion relation makes graphenes' lateral *p-n* junctions to be, in principle, highly transparent interfaces that support both conventional and negative electron refraction [13,14,18]. Not only that, the transmission probability $T_{pn}$ across graphene *p-n* junctions, which is theoretically determined by chiral (Klein) tunneling processes occurring at these interfaces, strongly depends on both the incident angle of the electron beam $\theta$ [19] and the width of these *p-n* junctions $w$ (Fig.1a) [20,21]. For a symmetric and bipolar *p-n* junction with a potential having a linear profile at the *p-n* interface, $T_{pn}$ at a specific electron Fermi wavelength $\lambda_F$ is given by the expression [14] $T_{pn} \sim e^{-2\pi^2 w/\lambda_F \sin^2\theta}$ (Fig 1.b). As a result, a broad range of novel electron-optics devices with different functionalities can be implemented in graphene by appropriate choices of *p-n* geometries (i.e $\theta$), and junction widths $w$. Such systems range from electron guiding devices [11,13-15] or beam collimators [22], to unconventional (Klein-tunneling) transistors employing the angle-dependent tunneling of Dirac electrons to engineer a gate-tunable current modulation in graphene [23-25]. Also, more sophisticated architectures such as two-dimensional Dirac fermion microscopes could be conceived by designing spatial arrangements of source-drain contacts and *p-n* junctions in graphene similar to the arrangement of electron sources, detectors and optics in an electron microscope [26,27]. The basic design principles for graphene's *p-n* interfaces are two [20-22]: *i)* sharp *p-n* junctions where $w$ is much smaller than $\lambda_F$ (case $w/\lambda_F \ll 1$) allow a relatively wide range of incoming angles to be transmitted. These highly transparent interfaces allow the realization of the so-called Veselago lenses, capable of focusing diverging trajectories of electrons emanating from a point source (Fig. 1c). *ii)* smooth *p-n* junctions (case $w/\lambda_F > 1$) represent interfaces where only electrons near normal incidence are transmitted while the rest are reflected, acting as electron beam (Klein) collimators (Fig. 1d). Considering typical carrier densities $n$ in graphene devices below $3 \times 10^{12}$ cm$^{-2}$ [11-15]; the condition of sharp *p-n* junction is fulfilled for widths $w \ll \lambda_F = 2\sqrt{\pi/n} \approx 20$ nm, i.e. for $w$ below 10 nm. For completeness, we note that sharp (and therefore smooth) lateral *p-n* junctions should additionally vary softly with respect to the lattice scale of graphene $a$ in electron-optics devices (i.e. $w \gg a \sim 0.2$ nm) to avoid inter-valley scattering in the system [21]. Thus, one can establish an approximated interval for useful, sharp lateral *p-n* junctions in graphene for junction widths $0.5 \leq w \leq 10$ nm.

*Implementing sharp lateral p-n junctions in graphene field-effect devices.*

From a technological perspective, sharp lateral *p-n* junctions are the key elements to achieve practical devices based on electron-optics principles. This is not only due to their aforementioned outstanding focusing ability, their high transparency enables a much larger carrier modulation (or contrast [22]) in these interfaces compared to smooth junctions [21, 22]; a vital aspect for applications. However, the technical implementation of sharp lateral *p-n* junctions in typical potential steps of the order of 0.1 eV [20-22] in graphene devices is highly challenging [13,14,24,25]. This is due to the prominent out-of-plane electric field present around lateral *p-n* interfaces, consequence of the system dimensionality; together with the weak screening of charge carriers existing in 2D materials due to their low density of states, [3,6,7] both of which considerably contributes to increasing the junction size $w$.

For instance, junction widths $w$ achieved via conventional local gating techniques such as bottom-, top- or split-gates are predominantly defined by the distance $b$ between the graphene sheet and the gates [13,14,28], which depends on the thickness of the dielectric material. The realization of sharper junctions could be attempted in these multiple-gate devices by using thin top and bottom dielectric layers with $b \leq 5$nm. Yet, in practice, these ultrathin dielectric films are limited by possible defects and/or ultimately quantum tunneling [29-31], compromising the proper functioning of the device.

In clear contrast, metal islands of different geometries deposited on graphene offer an alternative opportunity to realize sharp lateral *p-n* junctions in this 2D material (Fig.2). This is due to two main reasons: *i)* metals can dope graphene due to both electron transfer and chemical interactions taking place at metal-graphene interfaces [32,33], enabling the creation of lateral *p-n* junctions in the 2D material and *ii)* provide partial screening of the out-of-plane electric field existent at these *p-n* interfaces, reducing thus the junction width $w$. Indeed, sharp *p-n* junctions with potential steps of ~ 0.1 eV and $w \sim 1$ nm have been measured via scanning tunneling microscopy (STM) in continuous graphene sheets placed on copper [34,35], where the differently doped graphene regions occur at the interface of copper surfaces having different surface potentials. Not only that, it has also been possible to observe *gate-controlled* electron guiding in graphene by placing periodic arrays of metal nano-dots on graphene *field-effect devices* [15]. The observation of such guiding phenomenon in multiple (cascaded) *p-n* junctions at room temperature [15] relies on the formation of sharp *p-n* junctions at metal-graphene interfaces, even under the remote influence of the back-gate. The justification of this assumption has not yet been verified, nor has its implications been examined, and this is the main objective of this work. Whereas the existence of abrupt potential steps is generally assumed to occur at any metal-induced *p-n* interface in graphene field-effect devices (including contact regions) [15,36-40], realistic calculations of such potential profiles have never been carried out. These calculations are needed since the actual width *w* of the *p-n* junctions will depend on device geometries, dielectric environments and/or working conditions. In the present study, we undertake a systematic and robust investigation of the electrostatic problem of metal-induced lateral *p-n* junctions on graphene field-effect devices (i.e. accounting for the remote influence of the global back-gate) for electron-optics applications. Apart from the dependence on the carrier concentration *n* (i.e. Fermi level $E_F$) and height of the potential barrier $\Delta E$, four additional key factors could produce a noticeable impact on *w*: *i)* separation distance between metal and graphene $t_d$ ; *ii)* dielectric materials composing the device, including the possible presence of an ultra-thin dielectric at the actual metal-graphene interface; *iii)* the distance between the graphene sheet and back-gate *b* and, importantly, *iv)* device temperature *T* , due to the coexistence of thermally activated *n*- and *p*-type carriers at the interface. For any working condition (*n*, *T*), we find that lateral *p-n* junctions are generally sharper in devices where metal-graphene interfaces are embedded in environments with low dielectric constants $\varepsilon$ both above and below graphene, whereas distances *b* between graphene and back-gate do not significantly alter *w*. Temperature effects are more relevant for smaller potential barriers $\Delta E$, with *w* tending to decrease for higher *T*. Also, we discuss favorable device architectures in order to obtain sharper *p-n* junctions at room temperature (RT) and demonstrate that sharp junctions with *w*≤ 10 nm are indeed achieved at RT using non-encapsulated graphene supported on dielectric substrates with low permittivity such as SiO$_2$ or hexagonal boron nitride, hBN. Finally, we show how the precise details at the metal-graphene interface (such as separation distance $t_d$ and the dielectric constant of that interface $\varepsilon_d$) are critical to define the *p-n* junction in terms of both *w* and $\Delta E$. This is due to two simultaneous contributions to the total electric field occurring at these positions including the electric field present between metal and graphene layers responsible of creating the *p-n* junction; and the large out-of-plane electric field existing between these generated *p* and *n* regions. As demonstrated below, such interplay may promote the utilization of anisotropic thin dielectrics between metals and graphene to further control $w$.

We note that our results can be extended to other 2D materials since they are also doped by the presence of metal clusters [41,42] and are relevant to other applications where the spatial extent of lateral *p-n* junctions is important including graphene-metal contacts [36-39] and metal-graphene photodetectors [43,44]. The paper is organized as follows. In Section 2 the electrostatic model is presented. In Section 3 we show results of the junction width $w$ estimated in lateral *p-n* junctions created at metal-graphene interfaces for different device parameters such as gate voltage $V_g$, $\Delta E$, $\varepsilon$, $b$, $t_d, \varepsilon_d$ and *T*, and discuss device architectures promoting the formation of sharp *p-n* junctions at room temperature. Lastly, we present our conclusions in Section 4.

**2. Electrostatic model of *p-n* junctions at metal-graphene interfaces.**

When establishing lateral *p-n* junctions in 2D materials [3,6,7], a redistribution of charge occurs across the *p-n* interface to align the Fermi levels in both *p-* and *n-* regions while an in-plane potential $\phi$ with step height $\Delta\phi$ is established in the materials' plane ($z=0$) with a junction width $w$. Here, $\Delta\phi$ balances the difference between the intrinsic chemical potentials of the two regions, which are in thermal equilibrium. Solving the problem for a circular *p-n* junction with radius *R* in cylindrical coordinates $(r,\varphi,z)$ and having rotational symmetry along the azimuthal angle $\varphi$, the exact spatial distribution of the electrostatic potential $\phi(r,z)$ and thereby the width of the junction

$w$ can be obtained by solving the following non-linear 2D Poisson equation with appropriate boundary conditions (BC) in a convenient simulation region:

$$-\nabla \cdot [\varepsilon(r,z) \nabla \phi(r,z)] = \rho_{free}(\phi) \qquad \text{(Eq. 1)}.$$

It is important to emphasize that this expression of the Poisson equation is able to account for the presence of the back gate in the device through the free charge density $\rho_{free}$, the effect of the surrounding media via the local dielectric constant $\varepsilon$ and the screening produced by the metal island via convenient boundary conditions.

Three specific quantities have to be considered in the case of gate-tunable lateral $p$-$n$ junctions created at the interface between graphene regions underneath and outside a circular metal island in the device configuration shown in Fig. 2a. First, the generated in-plane electrostatic potential step $\Delta\phi$ should be equal to the energy shift $\Delta E \equiv -\Delta\mu = q\Delta\phi = \mu_g^{(m)} - \mu_g^{(0)}$ resulting from the charge redistribution at these one-dimensional interfaces [33,37,45], where $\mu_g^{(m)}$ and $\mu_g^{(0)}$ are the chemical potential in the graphene on the left ($L$) and right ($R$) regions, i.e. underneath and outside the metal (Fig.2b), respectively (situated far away from the $p$-$n$ junction interface, $r=R$) and $q$ is the elementary charge. In other words, there will be a negative or positive energy shift for $n$ and $p$ doping, respectively. Second, the free density charge across the junction is given by $\rho_{free} = \sigma(\phi)/t_G$ inside the graphene layer (zero outside), where $\sigma$ and $t_G$ are the surface charge density and the thickness of graphene, respectively. Here, the profile of the surface charge density $\sigma(r) = q(p(r) - n(r))$ is a consequence of both, the electrostatic gating due to the back gate potential $V_g$ tuning the overall Fermi level $E_F$ within the entire graphene device and the asymmetry of the chemical potential along the graphene sheet due to the presence of the metal island (Fig. 2b).

Importantly, $\sigma$ accounts for all mobile charges in graphene, electrons and holes with corresponding densities $n$ and $p$ respectively, coexisting across graphene's *p-n* interfaces at finite temperatures *T*. Third, the remaining device architecture is accounted for by the permittivity parameters $\varepsilon_1, \varepsilon_2$ (Fig. 2b). Unless otherwise stated, we consider a non-encapsulated graphene device (Fig. 2a) supported on a dielectric of relative permittivity $\varepsilon_2/\varepsilon_0 = 3.9$ (value corresponding to SiO$_2$ and/or hBN dielectric constants, commonly used in experiments [13-17]) and immersed in an environment of permittivity $\varepsilon_1$ equal to the vacuum permittivity $\varepsilon_0$. The separation distance from the graphene sheet to the back gate is $b$, effectively defining the thickness of the dielectric medium with permittivity $\varepsilon_2$ (Figs. 2a,b). The metal-graphene interface is represented [32,37,45] by a dipole layer formed as a result of the charge redistribution in the system within a common equilibrium separation distance $t_d$ = 0.3 nm [32,45] with a permittivity $\varepsilon_d = \varepsilon_0$. Additional details on how to obtain $\Delta\phi$ and numerically solve $\phi(r,z)$ with appropriate boundary conditions (BC) in a convenient simulation region are described in Supporting Information Notes 1- 4. We anticipate that *i)* graphene cannot be considered as a perfect metal due to its low carrier density (i.e. quantum capacitance effects should be considered, see Supporting Information Note 4) and *ii)* non-linear screening effects [28] beyond the Thomas-Fermi approximation [46,47] should be taken into account in the system since the characteristic lengths over which the potential varies across junctions formed at metal-graphene interfaces are smaller than $\lambda_F$. Also, *iii)* the actual potential of the metal island in this setup (Fig 2a) is constant but unknown, determined by the amount of charge existing in the island [33]. Besides, for completeness, *iv)* we note that our model is valid when the interaction between graphene and metals preserves the bandstructure of graphene [45]. Specific examples of metals and metal-graphene interfaces fulfilling this key condition are reported in Supporting Information Note 5.

Importantly, the possibility to solve Eq.1 without relying on restrictive hypothesis [7] and including (long-range) Coulomb interactions from both metal island and back gate [28,48] are key to simulate the electrostatics of metal-graphene interfaces in graphene field-effect devices, obtain accurate calculations of both potential profile $\phi(r)$ and/or charge distribution $\sigma(\phi(r))$ and extract critical quantitative parameters such as $w$ in the graphene sheet.

Fig. 2c shows both the simulated profile of the potential energy $-q\phi(r)$ and the energy shift $\Delta E \equiv -\Delta\mu = \mu_g^{(m)} - \mu_g^{(0)}$ within the graphene plane along a lateral $n$-$p$ junction, where the Fermi level is taken as reference ($E_F = 0$). The calculation is undertaken at room temperature ($T = 300$ K), using a metal that $n$ dopes graphene with $\Delta E = -0.2$ eV (close to values given by Ti or Ca [15,49,50], see Supporting Information Note 3) and considering the graphene sheet placed on a $b = 300$ nm thick layer of $SiO_2$. Furthermore, this calculation is done for a symmetrically doped (bipolar) $n$-$p$ junction at the specific back-gate voltage $V_g$ where the overall Fermi level $E_F = \Delta E/2$ (see Supporting Information, Note 3). In our study, the extraction of $w$ is done by tracing and extrapolating the slope of the curve $-q\phi(r)$ at the Fermi level up to the step $\Delta\mu = -q\Delta\phi$, as commonly done in literature [13]. Other reported ways to obtain $w$ are given by the width at which the 10% - 90% of the potential step is reached [14], which produces similar results.

The potential $\phi(r)$ mimics a Fermi function step with a width $w \sim 8$ nm, in agreement with the creation of sharp $p$-$n$ junctions ($w \leq 10$ nm) at metal-graphene interfaces in graphene field-effect devices. This width is considerably shorter than those reported by combinations of bottom-, top- and/or split-gates even when using thin dielectrics [12-14,24,25,38] ($w \geq 20$ nm in all cases). Variations of $w$ for different device parameters including $\Delta E$, $\varepsilon$, $b$, $T$, $t_d$, $\varepsilon_d$ and the presence of asymmetrically doped junctions at different gate-voltages $V_g$ (i.e carrier concentrations $n$) are discussed in the next section.

**3. Width of lateral p-n junctions at metal-graphene interfaces: back-gate voltage, potential barrier, device geometry and temperature dependences.**

Figure 3a shows the dependence of the junction width $w$ for different back-gate voltages $V_g$. All other device parameters (metal) and environmental conditions ($T$, $t_d, \varepsilon_d, b$) were similar to those of Fig. 2c. In general, these simulations correspond to asymmetric junctions in both bipolar $n$-$p$ (blue shade) and unipolar $p$-$p'$, $n$-$n'$ (gray shade) regimes with Fermi levels placed between, above or below the potential step $\Delta E$, respectively [21,39] (see Supporting Information Note 3). A decrement of $w$ from 12 nm to 5 nm is observed in the entire simulated range of $|V_g|$ from 0 to 50 V, respectively. From a physical point of view, the increase of $w$ when decreasing $n$ (or $|V_g|$) is a consequence of the low density of states (*DOS*) of graphene near the Dirac point [3]. In particular, one can analytically estimate (Supporting Information Note 4 and Ref. [3]) that $w$ should be proportional to $1/DOS \sim n^{-1/2}$, in good agreement with the trend shown in Fig. 3a.

Furthermore, as indicated in the figure by a continuous green line, these values of $w$ are well below the density-dependent Fermi wavelength $\lambda_F$ of the charge carriers in the device, thus the created lateral junctions are sharp for any $n$.

Fig. 3b shows $w$ for a symmetric dipolar $n$-$p$ junction ($E_F = \Delta E/2$, carrier concentration $n \sim 1\times10^{12}$ cm$^{-2}$) at $T$ = 300K for distances $b$ between the graphene sheet and the back gate varying from 10 to 400 nm. The latter values are commonly used in devices reported in literature [11-15,38,49]. Whereas it has some dependence on $b$, $w$ does not considerably increase within the calculated range for the here considered potential step $\Delta E = -0.2\text{V}$. All obtained values of $w$ are well below the corresponding $\lambda_F$ in the simulated case ~35 nm, and the created lateral junctions are therefore sharp in these cases. The absence of a major dependence of $w$ with respect to $b$ is ascribed to the effective screening of the out-of-plane electric field in the *n-p* junction by the metal island situated on top of (and extremely close to) the graphene layer.

Next, Fig. 3c displays $w$ for a symmetric dipolar *n-p* junction ($E_F = \Delta E / 2$, carrier concentration $n$ ~ $1\times10^{12}$ cm$^{-2}$) at $T = 300\text{K}$ and $b = 300$ nm for different relative dielectric constants of the supporting substrate $\varepsilon_2 / \varepsilon_0$. The simulated relative permittivity values $\varepsilon_2 / \varepsilon_0$ (1-100) are those present in common devices in literature [11-15], ranging from freestanding graphene ($\varepsilon_2 / \varepsilon_0 =1$), graphene supported on SiO$_2$ or hBN ($\varepsilon_2 / \varepsilon_0 =3.9$), or 'high-k' gate oxides such as HfO$_2$, $\varepsilon_2 / \varepsilon_0$ ~25 or TiO$_2$, $\varepsilon_2 / \varepsilon_0$ ~80. In general, $w$ is narrower for smaller $\varepsilon_2 / \varepsilon_0$, increasing notably when increasing $\varepsilon_2 / \varepsilon_0$, from 6 to 90 nm within the simulated range for the considered potential step $\Delta E = -0.2\text{V}$. Specifically, substrates with $\varepsilon_2 / \varepsilon_0$ below 10 show values of $w < 10$ nm (sharp junction case). In contrast 'high-k' dielectrics show widths $w \geq 50$ nm which are already larger than the corresponding $\lambda_F$ ~ 35 nm of the charge carriers. From a physical point of view, the increase of $w$ with use of media of higher permittivity can be understood from two linear contributions resulting from *i)* the dimensionality of lateral *p-n* junctions and *ii)* the fact that graphene is not a perfect metal. Specifically, *i)* surrounding dielectric media with higher permittivity makes the large out-of-plane electric field existing at lateral *p-n* junctions to vary more slowly, leading to a less efficient screening of such electric field (see Eq.1) and enlarging the width of the *p-n* junction. Besides, *ii)* quantum capacitance contributions may have to be considered at metal-graphene interfaces since graphene is not a perfect metal (as aforementioned in Fig.3a). In both cases, the length by which fields penetrate graphene at both sides of the junction shows a linear contribution to the permittivity of the surrounding medium (see Ref. [6] and Supporting Information Note 4 for further information about contributions *i)* and *ii)*, respectively). From a technological perspective, this result shows that the utility of 'high-k' dielectrics for electron optics applications might not be optimal (see detailed discussions below). Furthermore, we note that this observed dependence promotes the utilization of non-encapsulated over encapsulated devices to achieve sharper *p-n* junctions at metal-graphene interfaces. This is due to the fact that in encapsulated systems $\varepsilon_1 > \varepsilon_0$, increasing thus the overall effective dielectric constant of the device $\varepsilon_{eff} = (\varepsilon_1 + \varepsilon_2)/2$ and subsequently enlarging $w$, too [6]. Similar conclusions may apply in electron optics devices fabricated via conventional bottom-, top- or split- gates: here, suspended devices [12] would be preferred to achieve sharper *p-n* junctions over encapsulated ones [13,14].

Fig. 3d shows the temperature dependence of $w$ from 10K to 500 K for two symmetric dipolar lateral junctions ($E_F = \Delta E/2$) in a device with $b$ = 300 nm. The considered potential steps are $\Delta E_1$ = 0.08 eV and $\Delta E_2$ = -0.2 eV, corresponding to carrier concentrations at the junction $n_1 \sim 3 \times 10^{11}$ cm$^{-2}$ and $n_2 \sim 1 \times 10^{12}$ cm$^{-2}$ and Fermi wavelengths $\lambda_{F1}$ = 65 nm and $\lambda_{F2}$ = 35 nm, respectively. First, we can see that for any given temperature, $w$ is smaller for larger potential steps, in particular $w$ is approximately ~11 nm and ~8nm for $\Delta E_1$ and $\Delta E_2$, respectively. Furthermore, when sweeping *T*, we note that *i)* $w$ decreases for higher *T* in both cases $\Delta E_1$ and $\Delta E_2$ and *ii)* the overall variation of $w$ ($|\Delta w| = |(w_{max} - w_{min})/w_{max}|$) is larger for smaller steps $\Delta E$. Both interesting trends can be understood due to the influence of thermally activated carriers $n_{th}$ in the system [3]. In the first case, the larger number of $n_{th}$ at higher *T* will lead to a more efficient screening in the system [28] decreasing thus $w$, as commonly observed in conventional semiconductors [51]. Then, thermally activated carriers in a symmetric bias condition will have a larger impact in junctions with smaller step height, cases where the ratio $n_{th}/n_1$ is larger: one can see how $|\Delta w|$ decreases by 30% for $\Delta E_1$ when increasing the temperature up to 400K. Nevertheless, all calculated values of $w$ are well below the corresponding $\lambda_F$ in the three simulations, indicating that lateral p-n junctions remain sharp in the simulated temperature ranges and $\Delta E$.

Fig.3e shows the variation of $w$ with respect to the metal-graphene distance $t_d$ calculated for symmetric p-n junctions. Such distance depends not only on the metal type but also on other factors such as the equilibrium configuration [32]. In this case, $w$ increases when $t_d$ increases. Specifically, a considerable variation of $w$ from 7 to 12 nm is observed when increasing $t_d$ between 0.2 and 1.1 nm. This behaviour is qualitatively similar to the spread of the in-plane electric potential created in semiconductor field effect transistors with local gates, when these gates are separated away from the semiconductor material (see Refs. [52,53] and Supplementary Information Note 6). However, in our metal-graphene case, the potential step of the junction $|\Delta E|$ will also decrease when increasing $t_d$ since the energy shift of graphene underneath the metal depends on the actual graphene-metal distance. For instance, $|\Delta E| = 0.1$ eV when $t_d = 1.1$ nm, whereas $|\Delta E| = 0.2$ eV when $t_d = 0.3$ nm (with all other simulation conditions similar to those for Fig. 2c). As such, the increase of $w$ when increasing $t_d$ will not only be due to the device geometry (Supplementary Information Note 6), but also to quantum capacitance effects playing a role in the system (Supplementary Note 4). Explicitly, when increasing $t_d$, $|\Delta E|$ and the corresponding carrier concentration to obtain a symmetric dipolar lateral junction $n^*$ will decrease, too. The latter will also increase $w$, as already reported in Fig.3a. Regarding electron-optics applications, we note that all values of $w$ are well below the corresponding Fermi wavelengths $\lambda_F$ in the simulated cases, thus, created lateral junctions are sharp in all these cases ($t_d \leq 1.1$ nm).

Fig. 3f shows the dependence of $w$ with respect to the relative permittivity at the metal-graphene interface $\varepsilon_d/\varepsilon_0$. This parameter reflects, for instance, possible oxidation effects occurring in some metals at the interface with graphene [49]. Whereas an abrupt decrement of $w$ from 8.3 to 5.8 nm is observed when increasing $\varepsilon_d/\varepsilon_0$ from 1 to 10; $w$ decreases at a lower rate when $\varepsilon_d/\varepsilon_0$ is larger than 10. This overall behaviour can be intuitively understood from the fact that the electric field between metal and graphene (the one responsible of creating the actual *p-n* junction) will be more directional (perpendicular to the graphene film) in a medium with a higher permittivity (Eq.1). In this sense, the overall decrease of $w$ when increasing $\varepsilon_d/\varepsilon_0$ seems to follow a qualitative $(\varepsilon_d/\varepsilon_0)^{-1/2}$ dependence which, once again, is similar to trends occurring in depletion regions created by local gates in field effect transistors made from conventional semiconductors (see Refs. [52,53] and Supplementary Information Note 6). However, as in the previous case, we note that the potential step of the junction $|\Delta E|$ at metal-graphene interfaces additionally varies (increases) when increasing $\varepsilon_d/\varepsilon_0$. For instance, $|\Delta E| = 0.32$ eV when $\varepsilon_d/\varepsilon_0 = 80$, whereas $|\Delta E| = 0.2$ eV when $\varepsilon_d/\varepsilon_0 = 1$ (other simulation conditions are similar to the ones in Fig. 2c). Thus, the increase of $w$ when increasing $t_d$ will not only be due to the device architecture (Supplementary Information Note 6), but also to quantum capacitance effects playing a role in the system (Supplementary Note 4). Explicitly, when increasing $\varepsilon_d/\varepsilon_0$, $|\Delta E|$ and the corresponding carrier concentration to obtain a symmetric dipolar lateral junction $n^*$ increase, too. Such fact also decreases $w$, as already reported in Fig.3a. All values of $w$ are well below the corresponding Fermi wavelengths $\lambda_F$ in the simulated cases, and the created lateral junctions are therefore sharp in these cases.

Finally, we emphasize the interesting (opposite) trends exhibited by $w$ when changing the relative permittivities of the supporting substrate $\varepsilon_2/\varepsilon_0$ ($w$ increases when $\varepsilon_2/\varepsilon_0$ increases) and the dielectric existing at metal-graphene interfaces $\varepsilon_d/\varepsilon_0$ ($w$ decreases when $\varepsilon_d/\varepsilon_0$ increases). In general, the electric field between metal and graphene layers responsible of generating the *p-n* junction will be more directional (perpendicular to the graphene film) in a medium with a higher permittivity (thus, reducing *w*). Meanwhile, (by virtue of Eq.1) a lower permittivity in all surrounding dielectrics (not only $\varepsilon_2/\varepsilon_0$, but also $\varepsilon_d/\varepsilon_0$) guarantees the out-of-plane electric field existing between *p* and *n* sides of the junction to screen in a shorter distance within the graphene plane (thus, reducing *w*, too). This interesting interplay suggests the utilization of anisotropic thin dielectrics at metal-graphene interfaces ($\varepsilon_d/\varepsilon_0$) to custom tailor *w*. Specifically, to reduce further *w*, these dielectrics should have a large out-of-plane and low in-plane dielectric constants. An example is depicted in Fig. 4, showing the in-plane ($z = 0$) potential energy $-q\phi(r)$ between the graphene zone under the metal and the zone outside the metal around $r$ = R, where the thin dielectric used at the metal-graphene interface is anisotropic and has parallel and perpendicular dielectric constants 1 and 50, respectively. With these parameters, *w* = 4.4 nm, 45% smaller than the one shown in the isotropic case (*w* = 8nm, Fig 2c).

## 4. Conclusions.

In conclusion, we have performed a systematic and realistic study of the electrostatic problem of *p-n* junctions generated at metal-graphene interfaces. Junctions created at these interfaces can be sharp ( < 10 nm) and thus they can be potentially used in electron-optics applications. Apart from specific details of the metal-graphene interface (such as the separation distance between graphene and metal or the permittivity at this interface), the widths *w* of these lateral junctions considerably depend on the device architecture (i.e. dielectric environment) and experimental parameters including both carrier densities in the graphene sheet (i.e. different gate-voltages in the device) and temperature. Interestingly, our results promote the usage of device dielectrics with low permittivities, and higher temperature operation to reduce the width *w* of lateral *p-n* junctions created at metal-graphene interfaces in graphene field-effect devices. The technological relevance of metal-graphene interfaces in graphene electronics point towards the realization of systematic experimental studies to improve the understanding of these intriguing interfaces. In particular, several questions remain open in this field including the fine control of the separation distance metal-graphene; the perturbation of graphene's bandstructure due to the presence of metals [32,45] (see Supporting Information Note 5) including the possible creation of defects due to the metal deposition; the effect of metal granularity [54] in the dipole created at metal-graphene interfaces or undesired residual current being injected from the graphene to the metal in some cases (see Supporting Information Note 7). Finally, our results can be extended to other 2D materials [41,42] and are relevant to other applications where the spatial extent of lateral p-n junctions at metal-graphene interfaces is important, including contacts [36-39] and photodetectors [43,44].

**Acknowledgements** We acknowledge stimulating discussions with M. Brandbyge and G. Calogero. This work was supported by the Danish National Research Foundation Center for Nanostructured Graphene project DNRF103, DFF-EDGE (4184-00030) and the Union's Horizon 2020 research and innovation programme (grant agreement No GrapheneCore2 785219) and the Ministerio de Economía y Competitividad under the project TEC2015-67462-C2-1-R. J. E. S. acknowledges support by the European Structural and Investment Funds in the FEDER component, through the Operational Programme for Competitiveness and Internationalization (COMPETE 2020) [under the Project GNESIS - Graphenest's New Engineered System and its Implementation Solutions; Funding Reference: POCI-01-0247-FEDER-033566].

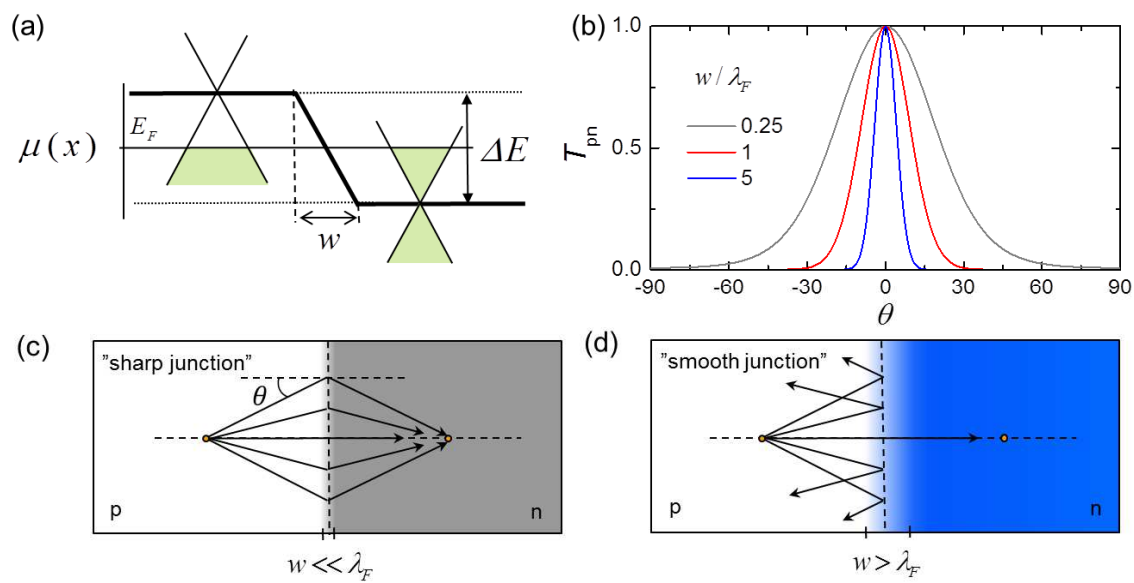

**Figure 1**. **Lateral *p-n* junctions in graphene** (a) Graphene bandstructure near a *p-n* junction with a potential step $\Delta E = -\Delta\mu = q\Delta\phi$ and transition width $w$. (b) Angle-selective transmission probability $T_{pn}(\theta)$ through a potential step strongly depends on the ratio $w/\lambda_F$. (c) Refraction at a sharp *p-n* junction with width $w/\lambda_F \ll 1$ exhibiting Veselago lensing. (d) Same as (c) for a smooth *p-n* junction with width $w/\lambda_F \gg 1$, leading to ray collimation: rays incident with large $\theta$ are specularly reflected.

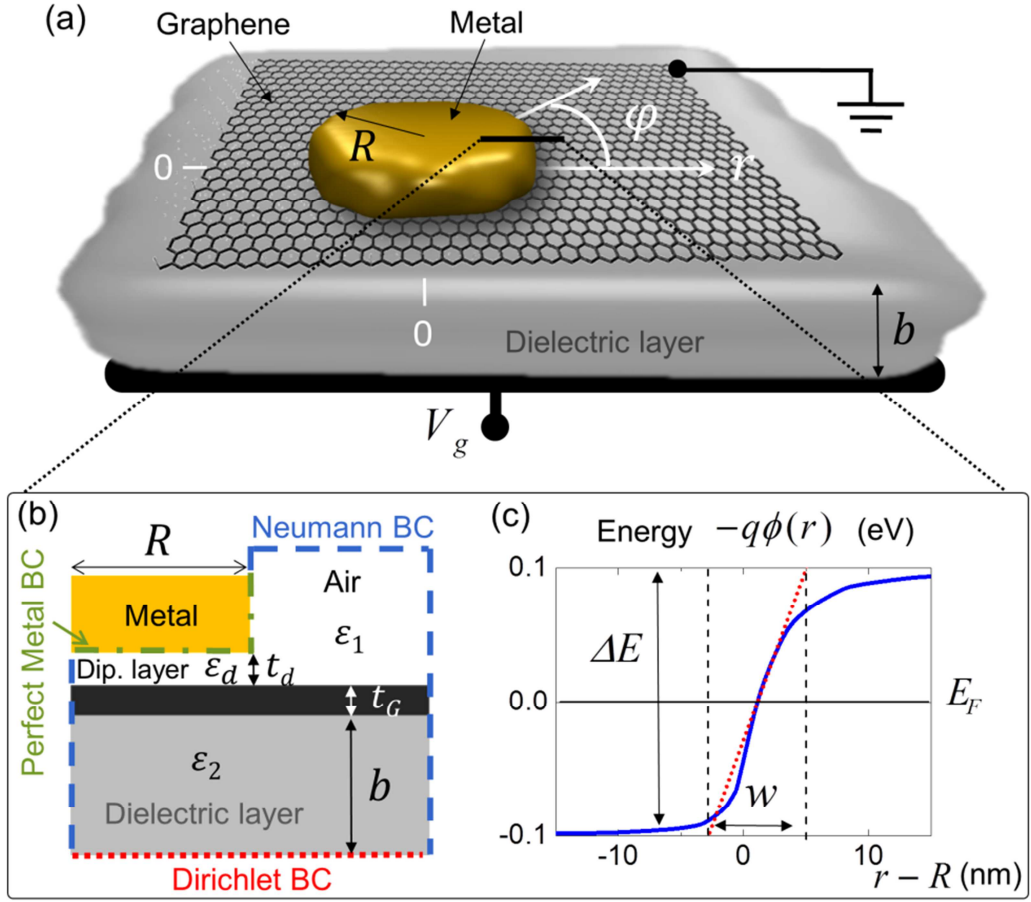

**Figure 2. Sharp lateral *p-n* junctions formed at metal-graphene interfaces.** (a) Schematic of a metal island deposited on a back gated graphene device. *p-n* junctions are formed in graphene at the edge of the metal island. (b) Sketch of the device along the angle $\varphi = 0$. Discontinuous lines represent the borders of the calculation region, fulfilling Neumann-like (blue dashes) and Dirichlet boundary conditions (BCs). Specifically, Dirichlet BC with a known potential is applied at the gate electrode (red dots), meanwhile a perfect metal BC is applied at the metal island (green dashed-dotted lines) (Supporting Information, Note 1). (c) In-plane ($z = 0$) potential energy $-q\phi(r)$ between the graphene zone under the metal and the zone outside the metal around $r = R$, with junction width $w$. Both $\phi(r)$ and $w$ are calculated by iteratively solving Eq.1. The simulation is done at room temperature in cylindrical coordinates around the point $(R,0,0)$ in order to simulate the circular metal island with radius $R = 50$ nm. We use Ti as the metal on top of graphene and assuming graphene placed on a 300 nm thick layer of $SiO_2$. We extract $w$ by the slope at the Fermi level (red dotted line) as commonly done in literature [13]. Our model reflects the fact that Ti *n* dopes graphene, in agreement with experiments [15, 49].

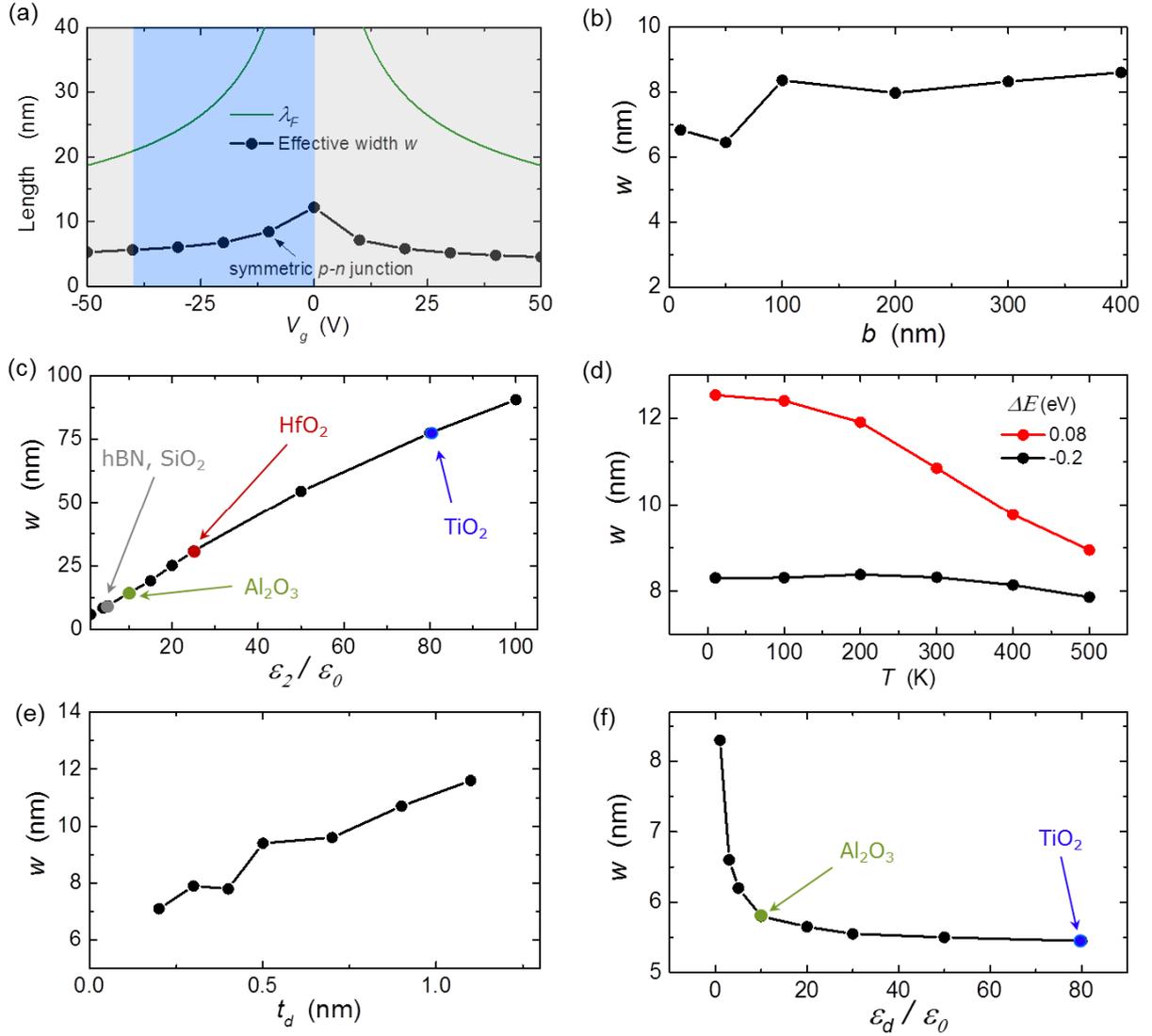

**Figure 3. Variation of the width $w$ of lateral $p$-$n$ junctions in graphene devices.** (a) Dependence of $w$ on the back-gate voltage $V_g$ at a constant $T$=300K and $b$ = 300 nm (Ti is used as metal). Light blue and gray regions correspond to bipolar ($p$-$n$, $n$-$p$) and unipolar ($n$-$n'$, $p$-$p'$) junctions, respectively. (b) Dependence of $w$ on the distance to the back-gate $b$ at $T$=300K for a potential step $\Delta E$ = -0.2 eV at $E_F = \Delta E/2$. (c) Dependence of $w$ on the dielectric constant of the supporting substrate $\varepsilon_2$ at $T$ = 300K and $b$ = 300 nm for a potential step $\Delta E$=-0.2eV at $E_F = \Delta E/2$. (d) Dependence of $w$ on the temperature $T$ ($b$ = 300 nm) for potential steps $\Delta E$ = 0.08eV (red points) and $\Delta E$ = -0.2eV (black points) at $E_F = \Delta E/2$. (e) Dependence of $w$ on the metal-graphene distance

$t_d$ at $T=300K$ at $E_F = \Delta E/2$. (f) Dependence of $w$ on the dielectric constant of the metal-graphene interface $\varepsilon_d$ at $T = 300K$ and $b = 300$ nm at $E_F = \Delta E/2$.

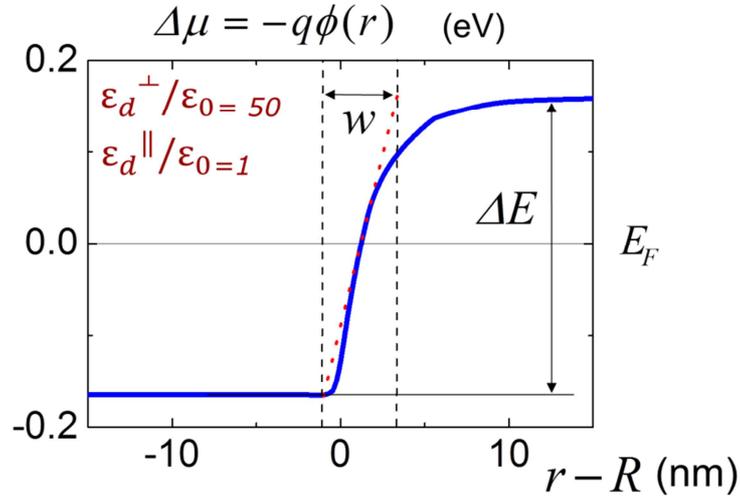

**Figure 4. Width $w$ of lateral p-n junctions with anisotropic dielectric constant at the metal-graphene interfaces.** In-plane ($z = 0$) potential energy $-q\phi(r)$ between the graphene zone under the metal and the zone outside the metal around $r = R$, with junction width $w$ for a metal-graphene interface with anisotropic parallel $\frac{\varepsilon_d^{\parallel}}{\varepsilon_0}=1$ and perpendicular $\frac{\varepsilon_d^{\perp}}{\varepsilon_0}= 50$ dielectric constants. In this case, $w = 4.4$ nm, 45% smaller than the one shown in the isotropic case ($w = 8$nm, Fig 2c).

# Supporting Information

# Electrostatics of metal-graphene interfaces: sharp *p-n* junctions for electron-optical applications

Ferney A. Chaves[1], David Jiménez[1], Jaime E. Santos[2], Peter Bøggild[3], José M. Caridad[3†]

[1]*Department d'Enginyeria Electrònica, Escola d'Enginyeria, Campus UAB, Bellaterra, 08193 Barcelona, Spain*

[2]*Centro de Física and Departamento de Física, Universidade do Minho, P-4710-057 Braga, Portugal*

[3]*Center for Nanostructured Graphene (CNG), Department of Physics, Technical University of Denmark, 2800 Kongens Lyngby, Denmark*

[†]corresponding author: jcar@dtu.dk

**Supporting Information Note 1. Numerical calculation of $\phi(r,z)$ and $\sigma(\phi)$ in the device.**

We solve the non-linear Poisson's equation, described by Eq. 1, in the *z=0* plane (i.e. the graphene plane) in cylindrical coordinates by a numerical method. The graphene sheet is grounded at a distance far away from the metal-graphene interface. Rotational symmetry is assumed along the azimuthal angle $\varphi$ and the graphene layer is located between *z* = 0 and *z* = $t_G$ (Fig.2). The borders of the calculation region fulfill Neumann-like condition $\frac{d\phi}{d\hat{n}}=0$ (vanishing electric field, blue dashed lines in Fig. 2b), where $\hat{n}$ is the direction normal to the border, except metal-like borders (red dashed and green dashed-dotted lines in Fig. 2b). Dirichlet-like conditions with a known potential, i.e. $\phi=V_g$ are imposed for the back-gate. Meanwhile, perfect metal (PM) boundary conditions are imposed in the metal island [S1]. This is due to the fact that the island is, in principle, at an unknown potential given by the actual charge of the cluster. The latter depends not only of the charge in the graphene layer, but also on the finite size of the island (i.e. size, shape and density of states, *DOS*) [S2]. For the present study, we consider *p-n* junctions where lengths of both *p* an *n* regions are well above the junction width *w*. This is, in our case such junctions are created by thin-film metallic islands on graphene (i.e. cylinders or stripes) with feature sizes (radius or width) > 10 nm. This condition (large cluster size) guarantees the *DOS* of the metal islands to be that one of the bulk metal, as demonstrated below in Note 2.

We further note that to avoid an erroneous calculation of the simulated out-of-plane field, electrostatic potential and junction widths due to Neumann boundary conditions [S3], our simulation region ($\pm r_{max}$, $\pm z_{max}$) is much larger than the calculated widths *w* [S3], at least by an order of magnitude. In particular (Fig. S1), ($\pm r_{max} = 250$ nm, $\pm z_{max} = t_s = 300$ nm).

The total free surface density $\sigma$ depends on the carrier densities in valance *p* band and conduction *n* bands:

$$\sigma = q(p-n). \text{ (Eq. S1)}$$

These surface densities *p* and *n* are deduced from the typical linear dispersion of graphene and the Fermi-Dirac thermal distribution and can be expressed in terms of the electrostatic potential $\phi$ as:

$$n = N_G F_1\left\{\frac{E_F - E_D}{k_B T}\right\} = N_G F_1\left\{\frac{-q\phi}{k_B T}\right\}$$

$$p = N_G F_1 \left\{ \frac{E_D - E_F}{k_B T} \right\} = N_G F_1 \left\{ \frac{q\phi}{k_B T} \right\} \quad \text{(Eqs. S2)}$$

Where $E_D$ is the Fermi energy at the Dirac point, $q$ is the elementary charge, $k_B$ is the Boltzmann's constant and $T$ is the absolute temperature of the device. $N_G = \frac{2}{\pi} \left( \frac{k_B T}{\hbar v_F} \right)^2$ is the density of states of the graphene sheet, thus, we take into account quantum capacitance effects in the system. These effects may cause [S4] lack of screening at these *p-n* interfaces where the quasiparticle density is very small (see Supplementary Information Note 4). Finally, $F_1(x) = \int_0^\infty \frac{u}{1 + e^{u-x}} du$ is the first order complete Fermi-Dirac integral. The solution of the 2D Poisson's equation with the corresponding boundary conditions is obtained by using an algorithm based on the Gauss-Newton iteration scheme applied to the finite element matrix coming from a finite element mesh.

The out of the plane equipotential lines for an exemplary situation corresponding to a symmetric *pn* junction are shown in the Fig. S1. Furthermore, we have considered that the graphene has a thickness of $t_G$ = 0.5 nm and an in-plane relative dielectric permittivity of $\varepsilon_G$ = 4 [S5, S6]. Also, the thickness $t_d$ and the relative dielectric permittivity of the dipole layer between the metal and the graphene $\varepsilon_d$, are chosen to be 0.3 nm and the vacuum permittivity $\varepsilon_0$, respectively [S7].

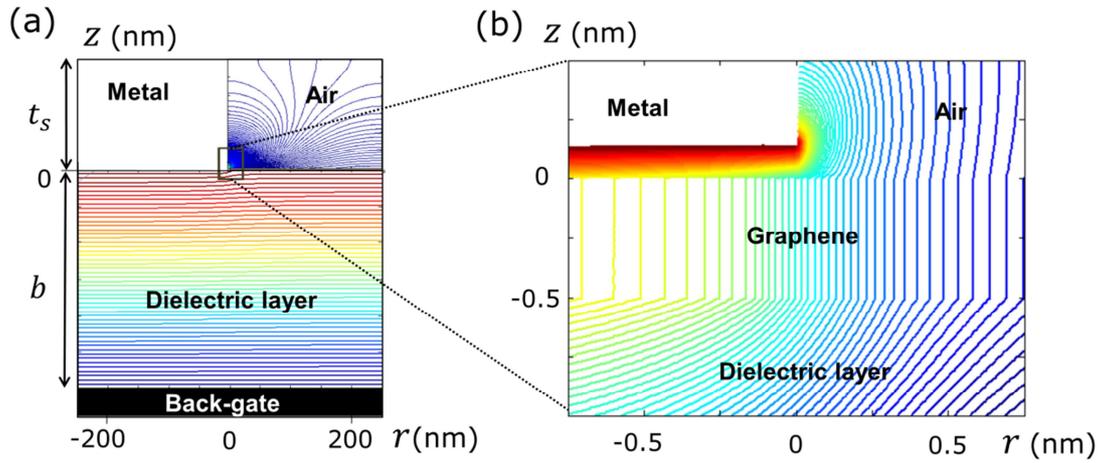

**Figure S1. Equipotential lines at metal-graphene interfaces.** (a) and (b) Equipotential lines in a graphene device at a metal-graphene interface and surrounding environment in the plane *r-z* (the sketch of the device is shown in Fig.2b).

Finally, we have double checked the validity of this electrostatic model for three cases:

*(i)* Confirming the values of $\Delta E = q\Delta\phi$ given by our model to the thermodynamic stability analysis proposed in Ref [S7]. In both cases $\Delta E \sim -0.2$ eV for a symmetric lateral *p-n* junction created in graphene by Ti as metal (see Fig. 2c, main text and Supporting Information Note 3).

*(ii)* Confirming that our electrostatic model gives similar junction widths *w* to the ones reported in literature in devices with multiple-gates, for instance, when a lateral *p-n* junction created in graphene encapsulated between hexagonal boron nitride (hBN) with a local top gate and a global bottom-gates in the device as done in Ref. [S8]. Incorporating the simulation parameters from Ref. [S8] in our electrostatic model, having a thickness of the top hBN = 15 nm and being the permittivity of this material 3.9, we obtain a *w* ~ 25nm, very close to the value reported in [S8] (~24nm).

*(iii)* Verifying our model with experiments. Scanning tunneling microscope (STM) is a high-resolution technique that can be employed to probe accurately the width of p-n junctions created at metal-graphene interfaces [S9]. Sharp *p-n* junctions with potential steps of the order of ~0.1 eV and $w \sim 1\text{-}3$ nm have been measured via scanning tunneling microscopy (STM) in continuous graphene sheets placed on copper [S9], where the differently doped graphene regions occur at the interface of copper surfaces having different surface potentials.

We have performed electrostatic simulations of a graphene sheet on two different metals, configuration showing potential steps of the order of ~0.1 eV, i.e. similar conditions to those reported in Ref. [S9]. Fig. S2 shows the simulated *p-n* junctions with widths ~ 2.8 nm, i.e. very similar to the width measured by STM.

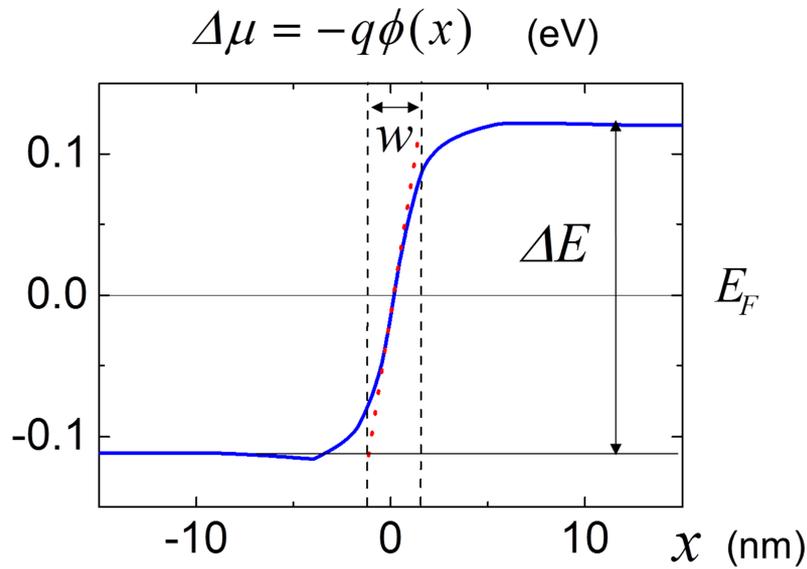

**Figure S2**. **Width *w* of lateral *p-n* junctions created in graphene supported by two metals with different surface potentials.** In-plane ($z = 0$) potential energy $-q\phi(x)$ in the graphene layer across the interface ($x = 0$) between the first and the second metal. The graphene-metal separation distances in this case are, $t_{d1} = 0.3$ nm and $t_{d2} = 0.5$ nm for metals 1 and 2, respectively, similar to experiments [S9]. In this case, the simulated $w$ = 2.8 nm, a value which is similar to the widths measured via STM [S9].

**Supporting Information Note 2. Effect of finite size on the density of states of metal islands.**

The level of doping of a graphene sheet in the presence of metal islands in a field effect transistor configuration depends not only on the bulk work functions of the different components of the system (metal,graphene) and the gate voltage applied to the device [S7], but also on a parameter that characterizes the chemical interaction between the graphene sheet and each individual cluster [S2]. Such parameter is subsequently determined by two contributions [S2]: the first one due to the induced surface dipole of graphene and of the metallic islands (together with other components of the system such as gate electrode); and the second one is the correction to the density of states of a cluster due to its finite size and specific shape. In the present study, we consider large metal clusters (i.e. metallic islands), where finite-size corrections to the bulk density of states $DOS(E_F^{Bulk})$ of the metal island at the Fermi level $E_F^M$ do not need to be introduced in our model. We show here that such corrections do not play a major role for metal islands larger than 10 nm. Specifically, considering a spherical metal cluster and using a free-electron gas approximation, finite-size corrections to the bulk density of states are given by [S2] : $DOS(E_F^M) = DOS(E_F^{Bulk}) - 3m^*/(8\pi\hbar^2 R)$, where $m^*$ is the effective mass of the metal atoms. Figure S3 depicts the $DOS(E_F^M)$ of an spherical Ti cluster depending on the radius $R$, showing how $DOS(E_F^M)$ and $DOS(E_F^{Bulk})$ display close values ( $DOS(E_F^M) > 0.96 * DOS(E_F^{Bulk})$ ) for spherical clusters with $R$ > 10 nm.

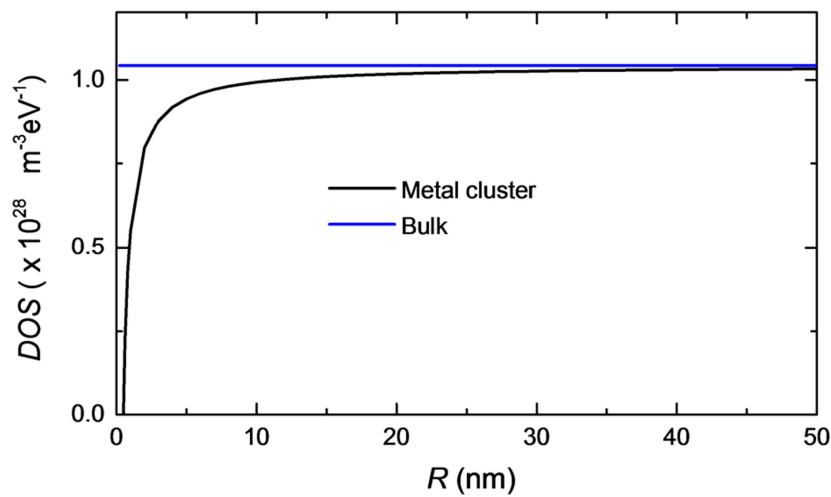

**Figure S3.** Corrections to the density of states (DOS) of a spherical metallic cluster made from Ti with radius $R$.

**Supporting Information Note 3. Estimation of $\Delta E$ from a thermodynamic stability analysis.**

In order to get a better understanding of the electrostatics of lateral *p-n* junctions at metal-graphene interfaces in graphene field-effect devices, we consider a thermodynamic stability analysis of the problem in the two graphene regions: underneath the metal and outside the metal, both of which need to be also in equilibrium with the overall back gate. The model is based on the imposition of thermodynamic equilibrium between the components of the system and allows us to estimate $\Delta E$ (or $\Delta\phi$) and to obtain the specific back-gate voltage $V_g$ where a symmetrically doped (bipolar) *p-n* junctions is established in the device (i.e. the condition where the Fermi level $E_F = \Delta E/2$). Before proceeding further, we note that graphene is modelled here as an infinite sheet, not considering finite-size effects such as inhomogeneous gating due to fringing electrostatic fields at the edges of the graphene sheet [S10]. Also, this model is exclusively valid away from the actual lateral junction. We then consider the two graphene regions underneath and outside the metal:

*Graphene region underneath the metal*

Fig. S4 shows the band diagram of the metal (M), dipole layer (DL), graphene (G), dielectric (D), back-gate (BG) vertical structure [S7,S11], where a *p-type* doped graphene has been assumed as a result of the metal-graphene interaction and the back-gate voltage, without loss of generality. Here $W_M$, $W_G$ and $W_{BG}$ are the metal, graphene and back-gate work functions, respectively, $\Delta V_{ox}$ is the voltage drop across the gate oxide, $\Delta V$ the voltage drop across the dipole layer formed between graphene and metal, $\mu_g^{(m)}$ is the Fermi energy variation of graphene underneath the metal, determined as a function of $V_G$ by solving the following set of Eqs. S3. These equations rise from the following conditions: *i)* the total charge density in the vertical heterostructure, including the metal surface charge density $Q_M$, the graphene layer surface charge density $Q_G$ and the back gate surface charge density $Q_{BG}$ must be zero (Eq.S3a) and *ii)* the sum of voltage drops around any loop (see Fig. S4) from the band diagram should be equal to zero (Eqs.S3b,c):

$$Q_M + Q_G + Q_{BG} = 0 \qquad \text{(Eq. S3a)}$$

$$W_M - q\Delta V - W_G - \mu_g^{(m)} = 0 \quad \text{(Eq. S3b)}$$

$$W_G + \mu_g^{(m)} + qV_G - q\Delta V_{ox} - W_{BG} = 0 \qquad \text{(Eq. S3c)}$$

As aforementioned, the graphene charge $\sigma$ below the metal is related to $\mu_g^{(m)} = E_F - E_D$ following the expression $Q_G(\mu_g^{(m)}) = q[p(\mu_g^{(m)}) - n(\mu_g^{(m)})]$. The surface charge densities $Q_M$ and $Q_{BG}$ are related to the voltage drop across the dipole and dielectric layers as $Q_M = -C_d \Delta V$ and $Q_{BG} = C_{ox}\Delta V_{ox}$, respectively; where $C_d = \varepsilon_d/t_d = \varepsilon_0/t_d$ and $C_{ox} = \varepsilon_2/b$ describe the dipole layer and back-gate capacitance per unit area. We note that, by performing this infinite parallel-plate capacitor approximation, clusters are assumed to be elongated objects. Combining Eqs. 3, we obtain the following transcendental equation for $\mu_g^{(m)}$ in graphene below the metal, which is numerically solved.

$$\frac{(C_d+C_{ox})}{q}\mu_g^{(m)} + Q_G(\mu_g^{(m)}) + \frac{C_d}{q}(W_G - W_M) + \frac{C_{ox}}{q}(W_G + qV_G - W_{BG}) = 0 \quad \text{(Eq. S4)}$$

Importantly, we note that this equation is valid for metal islands of any shape, as long as their density of states $DOS(E_F^M)$ is large (i.e. close to the one of the bulk metal). A more detailed analysis, including finite-size corrections to the bulk density of states $DOS(E_F^M)$ can be found in [S2].

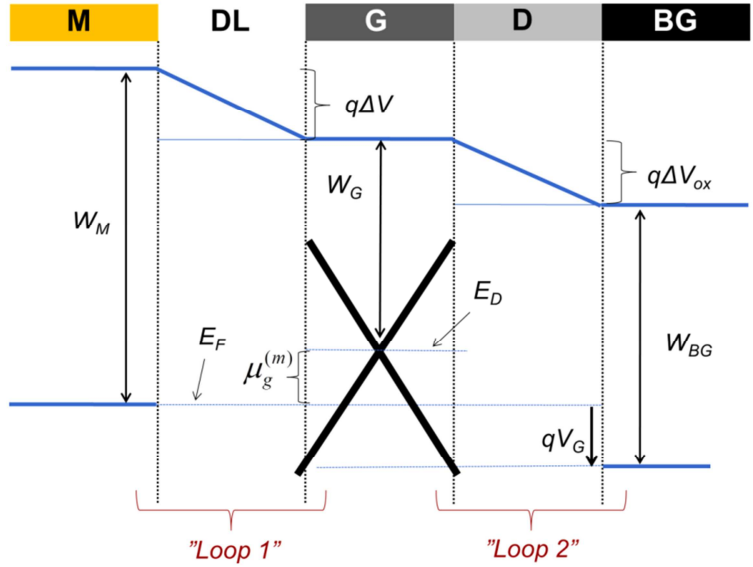

**Figure S4. Band diagram of a metal-graphene interface in a gated graphene device.** A voltage drop ΔV is produced over the dipole layer (Fig. 2b) and $\mu_g^{(m)}$ represents the shift of the graphene Fermi level $E_F$ with respect to the Dirac point $E_D$ due to both the metal presence and the back-gate.

*Graphene region outside the metal*

In a similar way, an equation for the shift of the graphene chemical potential outside of the metal region $\mu_g^{(0)}$ and far-away from the lateral junction can be obtained. Outside the metal zone Eq. S3a reduces to $Q_G + Q_{BG} = 0$, Eq S3b is not present and in Eq. S3c, $\mu_g^{(m)}$ is replaced by $\mu_g^{(0)}$. In other words, equilibrium is established between the back gate and graphene only.

Fig. S5 shows the calculated dependence of $\mu_g$ on the gate voltage $V_g$, in both graphene regions below $\mu_g^{(m)}$ and outside $\mu_g^{(0)}$ the metal, far-away from the lateral *p-n* junction. This is done considering titanium (Ti) as the metal, with the following overall parameters [S7,S12]: $W_G$ = 4.5 eV, $W_M$ = 4.33 eV, $W_{BG}$ = 4.5 eV, $\varepsilon_d$ = $\varepsilon_0$, $\varepsilon_2$ = 3.9$\varepsilon_0$, $T$ = 300 K and $b$ = 300 nm. By comparing both regions, we can see that Ti n dopes graphene and a symmetric *n-p* junction is created when a backgate voltage approximately equal to -10V is applied (condition $\mu_g^{(m)}(V_g) = -\mu_g^{(0)}(V_g)$). Also, the estimated step of the lateral *p-n* junction $\Delta E = -\Delta\mu = q\Delta\phi = \mu_g^{(m)} - \mu_g^{(0)}$ (see Fig.2, main text) is ~ -0.2 eV. Such value is a reasonable match to the -0.28 eV calculated from first principles [S12] and close to the values extracted from experiments, ranging from -0.12 eV [S13] to -0.15 eV [S14]. Furthermore, it is worth noting that values of $\Delta E$ from the 1D model accurately agree with those $|\Delta E| = |q\Delta\phi|$ obtained from the 2D model as aforementioned in section S1.

Finally, similar to the electrostatic model, this analysis is applicable to weakly bonded metals, those which do not change the bandstructure of graphene [S2,S7]. We note that, in practice, the strength of metal-graphene coupling (i.e the equilibrium distance metal-graphene) might be complex to determine, not only depending on the type of metal [S7], but also in the deposition conditions, metal granularity or annealing cycles [S14,S15]. Additional information about this can be found in the Note 5 of this Supporting Information.

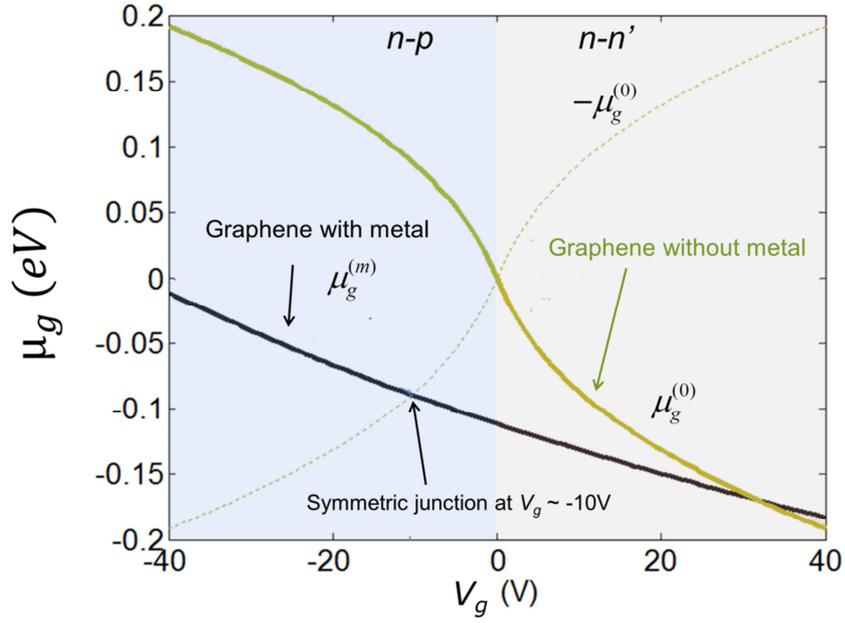

**Figure S5. Fermi level of graphene with respect to the Dirac point in regions with and without metal for different gate voltages $V_g$.** Ti is the metal selected for this figure, having a $W_M$ = 4.33 eV. A symmetric and bipolar lateral *n-p* junction is observed in this case for a gate voltage $V_g$ = -10 V.

**Supporting Information Note 4. Criterion for treating graphene as a perfect metal at metal-graphene interfaces**

In this section, we justify the fact that graphene cannot be considered as a perfect metal to calculate the width of *p-n* junctions *w* at metal-graphene interfaces.

We start from the net current density $j_e$ in a conductor, given by the expression [S16]:

$$j_e = \sigma E - \alpha \nabla \mu \qquad \text{(Eq. S5)}$$

where $\alpha \nabla \mu$ represents the diffusion current. In equilibrium $E = -\nabla \phi$ and $j_e = 0$, $\sigma/\alpha = q$, and we have:

$$\mu + q\phi = const. \qquad \text{(Eq. S5)}$$

At a constant temperature, and if the density of carriers $n$ is close to equilibrium (i.e. $\delta n = n - n^{eq} \ll 1$), $\nabla \mu = \frac{\partial \mu}{\partial n} \nabla \delta n$ and we can write:

$$j_e = \sigma E - \alpha \frac{\partial \mu}{\partial n} \nabla \delta n \qquad \text{(Eq. S6)}$$

Then, in equilibrium, and using Ficks law $\alpha \frac{\partial \mu}{\partial n} = qD$

$$\nabla \phi + \frac{qD}{\sigma} \nabla \delta n = 0 \qquad \text{(Eq. S7)}$$

where $D$ is the diffusion coefficient.

From $\sigma/\alpha = q$ and the fact that $\frac{\partial n}{\partial \mu} \approx DOS(E_F^g)$, $D = \sigma/\left(q^2 DOS(E_F^g)\right)$, which is the Einstein relation [S16]. Hence, using $\delta n = n - n^{eq}$, at the surface of the graphene:

$$\phi + \frac{1}{qDOS(E_F^g)} n = const. \text{ (Eq. S8a)} \quad \text{or} \quad \phi + \frac{\sigma_g}{q^2 DOS(E_F^g)} = const. \text{ (Eq. S8b)}$$

where $\sigma_g = qn$ is the surface charge density in graphene.

On the other hand, $\sigma_g$ is given by:

$$\sigma_g = -\varepsilon_1 \left.\frac{\partial \phi}{\partial z}\right|_{0^+} + \varepsilon_2 \left.\frac{\partial \phi}{\partial z}\right|_{0^-} \qquad \text{(Eq. S9)},$$

assuming that graphene is placed at $z=0$ and the dielectric constant above and below graphene are $\varepsilon_1 = \varepsilon_{1r}\varepsilon_0$ and $\varepsilon_2 = \varepsilon_{2r}\varepsilon_0$, respectively. For simplicity we take here $\varepsilon_1 = \varepsilon_2 = \varepsilon = \varepsilon_r \varepsilon_0$.

Using Eqs. S8 and S9, we have:

$$\left.\frac{\partial \phi}{\partial z}\right|_{0^+} - \left.\frac{\partial \phi}{\partial z}\right|_{0^-} - \frac{q^2 DOS(E_F^g)}{\varepsilon} \phi = const. \qquad \text{(Eq. S10)}.$$

In Eq. S10, one can define the carrier density dependent quantity

$$l_g = \frac{\varepsilon}{q^2 DOS(E_F^g)} \quad \text{(Eq. S11)}$$

which has dimensions of length. $l_g$ represents the scale at which the perfect metal approximation can be used for graphene. In other words, when $l_g = 0$ or, more generally, when $l_g$ is much smaller than any other geometrical lengths in the device, graphene can be considered as perfect metal, and the junction width $w$ will be entirely determined by geometrical factors. Otherwise, quantum capacitance effects need to be taken into account in the system due to the lack of screening at *p-n* interfaces [S4]. Here, $w$ will be larger, proportional to $l_g$ and dependent on additional parameters including device parameters such as the back gate dielectric and operational parameters such as temperature or carrier density.

In particular, given the density of states of graphene [S17] $DOS(E_F^g) = \frac{2|E_F|}{\pi \hbar^2 v_F^2} = \frac{2\sqrt{\pi n}}{\pi \hbar v_F}$, one can estimate $l_g$ at typical carrier densities of graphene $n=10^{12}$ cm$^{-2}$ to be ~ 0.5 nm when graphene immersed in vacuum $\varepsilon = \varepsilon_0$. As such, first, we cannot use the perfect metal approximation in metal-graphene interfaces, systems where the separation between graphene and metal (~0.3 nm [S7]) is comparable to $l_g$. Furthermore, we note that $l_g$ not only decreases at lower carrier densities $n$ proportionally to $l_g \sim n^{-1/2}$ but also linearly with the permittivity of the surrounding medium $\varepsilon$. This means that the screening of the in-plane electric field at the junction is less effective when increasing the permittivity of the surrounding media [S4], and is one of the reasons why $w$ increases when graphene is supported on 'high-k dielectrics' with respect to dielectrics with much lower permittivity (Fig. 3c, main text). We emphasize that the calculated $w$ (Figs. 3a and 3c, main text) follows pretty accurately the same trends than $l_g$: linear with respect to both $n^{-1/2}$ (Fig. 3a) and $\varepsilon$ (Fig. 3c).

**Supporting Information Note 5. Applicability of our model**

Our model is applicable when the interaction between metals and graphene is weak, i.e. the graphene bands, including their conical points at K, are preserved and can be clearly identified. Specifically, this situation occurs when the separation distance between graphene and metal is $t_d > 0.3$ nm [S7]. Graphene on metals such as Al, Cu, Ag, Au, Pt, show separation distances $t_d > 0.3$ nm [S7]. Not only that, such situations ($t_d > 0.3$ nm) also occur for certain configurations of graphene on alternative metals, despite such metals might be commonly regarded to interact strongly with graphene. For instance, this is the case of the so-called "BC" bonding configuration of graphene on Ni [S18].

In addition, we note that the determination of the actual equilibrium distance between metal and graphene $t_d$ (i.e. the strength of the metal-graphene coupling) might be more complex in practice. This will not only depend on the type of metal or equilibrium configuration, but also on device-specific conditions such as vacuum levels when depositing the metal on graphene, metal granularity, performed annealing cycles and/or the formation of a native oxide layer at the interface with graphene in some metals such as Al or Ti [S14, S19, S20]. More generally, we note that even in the case of graphene interacting strongly with some metals, the monolayer could be decoupled from the metallic substrate using different techniques (oxidation, intercalation of different atomic species or others [S21, S22]). All of these comprise examples where graphene's bandstructure will be preserved and thus our model will be applicable.

**Supporting Information Note 6. Dependence of the in-plane potential on $t_d$ and $\varepsilon_d$**

It is possible to analytically estimate the dependence of the in-plane potential $\phi$ on the separation distance between metal and graphene $t_d$ and the permittivity of this gap $\varepsilon_d$. Such problem is similar to the calculation of depletion lengths in locally gated field effect transistors made from thin semiconductor films (silicon on insulator, SOI) in order to avoid short-channel effects [S23,S24].

Here, one can show that a natural length $\lambda$ controls the spread of the potential distribution of $\phi(x)$ in the graphene plane along the $x$ direction (for convenience we use here Cartesian coordinates). Indeed, assuming a simple parabolic form of the potential distribution $\phi(x,z) = c_0(x) + c_1(x)z + c_2(x)z^2$ one can solve the Poisson's equation (Eq.1 main text) with the three boundary conditions of the problem [S23,S24]:

1- $\phi(x,0) = c_0(x)$

2- The electric field at the top of the graphene surface (z = 0, Fig S1b) is determined by the difference between the potentials at metal $\phi_M$ and graphene $\phi(x,0)$ surfaces and $t_d$ as:

$$\left.\frac{d\phi(x,z)}{dz}\right|_{z=0} = \frac{\varepsilon_d}{\varepsilon_g} \frac{\phi(x,0) - \phi_M}{t_d} = c_1(x)$$

3- The electric field at the bottom of the graphene surface (z = -$t_G$) is close to zero, giving: $c_1(x) - 2t_G c_2(x) \simeq 0$. We note that this approximation is valid when considering a weak field at the supporting dielectric substrate (i.e. no significant backgate potential).

Considering these boundary conditions and a constant graphene permittivity $\varepsilon_g$, Eq.1 can be written as:

$$\nabla^2 \phi(x,z) = \frac{\rho_{free}(\phi)}{\varepsilon_g} = \frac{d^2\phi(x,0)}{dx^2} + \frac{\varepsilon_d}{\varepsilon_g} \frac{\phi(x,0) - \phi_M}{t_G t_d} \quad \text{(Eq. 12)}$$

This equation can be solved [S23,S24] by undertaking the transformation $\lambda = \sqrt{\frac{\varepsilon_g}{\varepsilon_d} t_G t_d}$. Indeed, this parameter has units of length and effectively describes the potential distribution $\phi(x)$ of the interface. Without the need to solve the equation, we can clearly see how $\lambda$ increases when increasing $t_d$ and is proportional to $\varepsilon_d^{-1/2}$. These two trends are observed in our simulations in Figs. 3e and 3f, main text, respectively.

**Supporting Information Note 7. Preventing current injection from graphene to metal islands**

For a proper functioning of electron-optics devices such as Klein tunneling transistors, current injection from graphene to metal islands should be avoided. This is needed to achieve a large current modulation in these devices.

We undertake a simple resistor circuit analysis (Fig. S6) to quantitatively evaluate the possibility of injecting current from graphene to the floating metal island. As we will see, this calculation incorporates additional device parameters such as metal island size or graphene quality. This information is useful to understand further and design metal-graphene interfaces for electron-optics applications.

Here, we assume for simplicity that *i)* the sheet resistance of graphene underneath the metal $R_g^M$ is similar to the sheet resistance of the graphene without metal on top $R_g^0$: $R_g^M = R_g^0 = R_g^S$ (i.e. we solve the symmetric junction case). Furthermore, in this case, *ii)* the metal on top of graphene is a rectangle, width $W = 1$ µm and length $L_M$, dimensions which are larger than the current injection/ejection region $L_i$ from graphene to metal and viceversa ($L_i = \sqrt{\rho_C / R_g^S}$ where $\rho_C$ is the contact resistivity at the metal-graphene interface. [S25].

Based on this simple circuit, the current flowing through graphene underneath the metal $I_g$ with respect to the incident current $I_{in}$ is $I_g / I_{in} = 1/(1 + R_g / 2R_c)$ and takes values 84% and 96.2% for $L_M$ = 0.5 µm and $L_M$ = 100 nm, respectively. The former calculations have been performed assuming a graphene sheet with mobility µ = 20000 cm$^2$V$^{-1}$s$^{-1}$ at a carrier density $n = 2 \times 10^{12}$ cm$^{-2}$ ($R_g^S = (ne\mu)^{-1} \sim 150\Omega$), and typical contact resistivities at metal-graphene interfaces $\rho_c \sim 1 \times 10^{-5}$ $\Omega$cm$^2$ [S11, S26]. For both calculated cases $L_i$ is limited by $L_M$; and the graphene and contact resistances are given by the expressions $R_g = R_g^S L_M / W$ and $R_c = \sqrt{\rho_c R_g^S} \coth(L_M / L_i) / W \approx \sqrt{\rho_c R_g^S} / W$.

We note that these values are consistent with values calculated using a more advanced resistor networks model [S27], where > 75% of current flowing through graphene with a lower mobility µ = 5000 cm$^2$V$^{-1}$s$^{-1}$ for a length $L_M$ = 1 µm. In general, the larger $\rho_c$ and the shorter $L_M$, the higher percentage of current flows through graphene. As such, this ratio can be controlled and custom tailored by increasing the distance between metal and graphene $t_d$ (using the aforementioned techniques described in Supporting Note 5) since $\rho_c$ increases exponentially when increasing $t_d$, see Ref. [S26].

Finally, we note that the estimated current ratio values are consistent with experiments [S13] demonstrating the creation of electron optics devices by depositing metallic dots with sizes ~100 nm on graphene transistors with mobilities µ ~ 20000 cm$^2$V$^{-1}$s$^{-1}$.

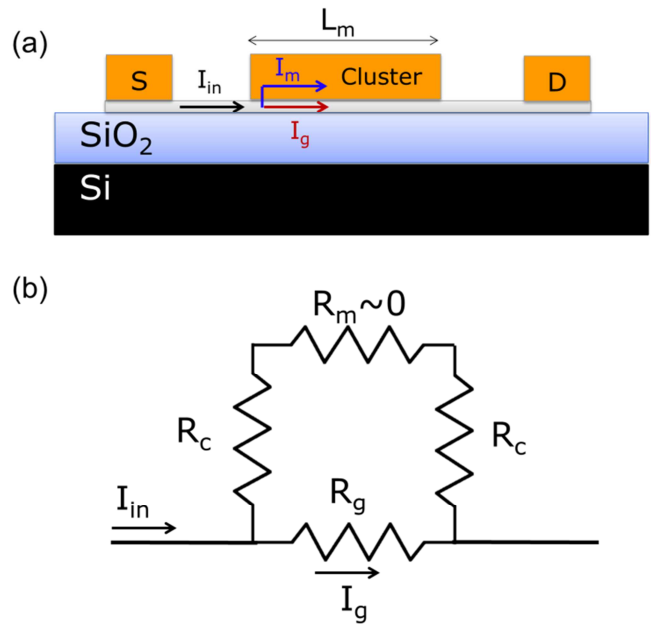

**Figure S6.** (a) Schematic of the device with the floating metal cluster of length $L_m$ on the graphene channel. (b) Equivalent circuit model, approximating the metal resistance $R_m$ to zero.